\documentclass[a4paper,onecolumn,11pt]{quantumarticle}
\pdfoutput=1
\usepackage[utf8]{inputenc}
\usepackage[english]{babel}
\usepackage[T1]{fontenc}
\usepackage{amsmath}
\usepackage{hyperref}
\usepackage{tikz}
\usepackage{lipsum}
\usepackage{bbold}
\usepackage{color}
\usepackage{amsmath}
\usepackage{amssymb}
\usepackage{amsthm}
\usepackage{comment}
\usepackage{braket}
\usepackage{graphicx}
\usepackage{breqn}
\usepackage{subfig}
\usepackage{lipsum}
\usepackage[backend=bibtex8,                
maxnames=6,                         
bibstyle=numeric,               
hyperref=true,                  
citestyle=numeric-comp, 
sorting=none]{biblatex}
\usepackage{lscape}
\DeclareMathOperator{\Tr}{Tr}

\newcommand{\euler}{e}
\newcommand{\ramuno}{i}
\newcommand{\qudit}[1]{\left\vert #1 \right\rangle}
\newcommand{\rqudit}[1]{\left\langle #1 \right\vert}

\newcommand{\Z}{\mathbb{Z}}
\newcommand{\id}{\mathbb{1}}
\newcommand{\C}{\mathbb{C}}

\newtheorem{thm}{Theorem}

\newtheorem{definition}[thm]{Definition}

\bibliography{mybibliography}

\begin{document}

\title{Toy model for the correlation of qudit bipartite states with maximally mixed marginals}

\author{Constantino Rodriguez-Ramos}\email{tinorodram@gmail.com}
\author{Colin M.~Wilmott}
\affiliation{Department of Mathematics, Nottingham Trent University, Clifton Campus, Nottingham NG11 8NS, UK}

\maketitle

\begin{abstract}
In this paper, we consider the local unitary classification of the class of qudit bipartite mixed states for which no information can be obtained locally. These states are represented by symmetrical density matrices in which both tracial states are maximally mixed. Interestingly, this symmetry facilitates the local unitary classification of two-qubit states. However, the same formalism fails in the case of systems of higher dimensions. We consider a broader set of states by introducing a family of qudit bipartite mixed states with maximally mixed marginals. For this family of states, we determine several constants which are in variant under local unitary transformations so can be used for entanglement classification. Finally, we consider the two-qutrit case and in particular, a two-parameter family of states for which the local unitary classification is complete. We relate this classification to known entanglement measures such as purity and negativity.
\end{abstract}

\section{Introduction}

Entanglement describes a particular type of correlation unique to composite quantum systems \cite{nielsen2002quantum,horodecki2009quantum}. One property of entangled systems is that the state of the system cannot be described by knowing the state of its constituent subsystems. Since the development of quantum information, entanglement was identified as a valuable resource in different scenarios. For example, entanglement is a crucial resource in applications such as teleportation \cite{bennett1993teleporting}, dense coding \cite{bennett1992communication}, quantum cryptography \cite{ekert1992quantum}, quantum computing \cite{ekert1992quantum}, and more recently in applications such as quantum sensing \cite{chalopin2018quantum} and quantum internet \cite{kimble2008quantum}. For this reason, tremendous efforts have been put into characterising the entanglement of quantum systems.

The simplest form of entanglement is bipartite entanglement. In this setup, the system is constituted by two parties, usually denoted as Alice and Bob. 
A unitary operation acting locally, only on Alice or only on Bob, preserves the entanglement of the joint system \cite{vedral1997quantifying,wootters1998entanglement}.
Establishing a complete set of local unitary (LU) equivalence classes is usually the first approach to characterise a bipartite system in terms of its entanglement. However, the complete LU classification of a general quantum system is usually difficult, and only partial results exist. For pure bipartite states, the complete LU classification is given by the set of Schmidt coefficients of the quantum state which can be computed by evaluating the spectrum of the marginals of the original state \cite{nielsen2002quantum}. For mixed states, establishing the LU classification of a general bipartite state is still an open problem. However, some remarkable contributions were made in this direction. In \cite{makhlin2002nonlocal}, a complete characterization of the LU classes of the two-qubit system is obtained in terms of 18 parameters. In \cite{sun2017local}, a complete set of invariant scalars of the LU classes of arbitrary dimension systems was obtained in \cite{zhou2012local}. However, this classification cannot be performed over states which are not full-rank. Zhang et al.\ presented a criterion to discriminate states that are not LU equivalent which is based on realignment and partial transposition of the states \cite{zhang2013criterion}.

In this paper, we investigate the LU classification of bipartite states with maximally mixed marginals. To do this, we adapt the formalism introduced in \cite{ramos2021convex} for quantum channel construction. This formalism allows us to consider a parameterized family of bipartite states with fixed rank. To establish LU classification for this family of states we evaluate sets of scalars that are invariant under local unitary transformations in terms of the defining parameters of the family. We consider the particular bipartite system composed of two qutrits. For this quantum system, we evaluate explicitly the sets LU invariant scalars in terms of the defining parameters of the family. This allows us to obtain analytic expressions for known quantum state measures such as purity and negativity.

The outline of this paper is as follows. In section \ref{Prelim}, we introduce the tools required for LU classification of bipartite states. In particular, we employ LU invariant scalars derived from the spectrum of the partially transposed matrix and the correlation matrix. In section \ref{BipartiteFamily}, we provide the parameter family of quantum states generalizing the anti-symmetric Werner state and maximally entangled pure states. In section \ref{LUinvariants}, we provide the explicit evaluation of the sets of LU invariant scalars. We find that for this particular family, the computation of the spectrum matrices is simpler than for a generic state. In section \ref{QutritCase}, we consider the qutrit-qutrit set-up. As an example, we show that for a bi-parametric family of states we evaluate explicitly the invariant scalars and we associate them to two known quantum measures of the states, purity and negativity.

\section{Preliminaries}\label{Prelim}
\subsection{Quantum states}

A $d$-dimensional quantum system can be described by pure states $\ket{\psi}\in\mathcal{H}_d$ where $\mathcal{H}_d$ corresponds to the Hilbert space endowed with the inner product $\braket{\psi_1|\psi_2}\in\mathbb{C}$ which is linear in the first argument and anti-linear in the second. We represent by $\mathcal{B}(\mathcal{H}_d)$ to the set of bounded linear operators acting on the pure states. Probabilistic mixtures of quantum states are given by the class of bounded linear operators called density operators $\rho \in \mathcal{B}(\mathcal{H}_d)$. In particular, these density operators correspond to positive semi-definite operators with unit trace, we denote the set of density operators by $\mathcal{D}_n$. A pure bipartite state $\ket{\psi^{AB}}$ is separable if it can be expressed as a tensor product of pure states, $\ket{\psi^{AB}}=\ket{\psi^{A}}\otimes\ket{\psi^{B}}$. In the case that a pure bipartite state cannot be expressed in as a tensor product of pure states we say that the state is entangled. We have that $\kappa\in\C$ is an local unitary (LU) invariant of $\rho_{AB}\in\mathcal{D}_d$ if
\begin{equation}
\kappa(\rho_{AB})=\kappa(\rho'_{AB})
\end{equation}
where
\begin{equation}
\rho'_{AB}=(U\otimes V)\rho_{AB}(U\otimes V)^{\dagger}
\label{localunitaryrotation}
\end{equation}
and $U,V\in SU(d)$ where $SU(d)$ correspond to the set of $d$ dimensional special unitary matrices.

\subsection{Quantum channels}

Consider a pair of quantum systems represented by the Hilbert spaces $\mathcal{H}_A$ and $\mathcal{H}_B$. A quantum channel from system $A$ to system $B$, denoted by $\mathcal{E}$ represents a linear mapping between the state spaces of both systems $\mathcal{E}:\mathcal{B}(\mathcal{H}_A) \rightarrow \mathcal{B}(\mathcal{H}_B)$. While $\mathcal{H}_A$  and $\mathcal{H}_B$ may have different dimensions, let us consider quantum channels between systems with the same size $\text{dim(}\mathcal{H}_A\text{)}=\text{dim(}\mathcal{H}_B\text{)}=d$. Quantum channels can be used to describe the evolution of a quantum system if we impose two extra conditions on their operator representation. These conditions are complete positivity and trace preservation. A quantum channel $\mathcal{E}$ is positive for any positive semidefinite operator $X$, $\mathcal{E}(X)$ is also positive semidefinite. Furthermore, $\mathcal{E}$ is completely positive if $\mathcal{E}\otimes\mathbb{1}_d$ is positive for all possible $d\in\mathbb{N}$ where $\mathbb{1}_d$ represents the identity operator acting on $\mathcal{H}_d$. The second condition is trace preservation which corresponds to the condition, $\Tr(\mathcal{E}(X))=\Tr(X)\:\forall\:X\in\mathcal{D}_d$. Finally, we may consider the particular class of unital quantum channels which are those channels mapping the maximally-mixed state to itself $\mathcal{E}(\mathbb{I}_d)=\mathbb{I}_d$. 

Consider the states of the joint system of $A$ and $B$ which are represented by bipartite states $\rho_{AB}\in\mathcal{B}(\mathcal{H}_A\otimes\mathcal{H}_B)$.  Despite the apparent differences, the mathematical representations of bipartite systems and quantum channels can be related through the Choi-Jamilkowski isomorphism \cite{jamiolkowski1972linear}. In particular, the Choi-Jamilkowski isomorphism establishes that, for every quantum channel $\mathcal{E}:\mathcal{B}(\mathcal{H}_A) \rightarrow \mathcal{B}(\mathcal{H}_B)$, we can associate a bipartite state given by \begin{equation}\label{CJ}\rho_{\mathcal{E}}=(\mathbb{I}_d\otimes \mathcal{E})(\ket{\psi}\bra{\psi}),\end{equation} 
where $\ket{\psi}=\frac{1}{\sqrt{d}}\sum_{j=1}^d\ket{j}\ket{j}$ is a maximally entangled state of dimension $d$. Using the relation given by \eqref{CJ}, we obtain that unital quantum channels correspond to bipartite states with maximally mixed partial traces
\begin{equation}
\Tr_A(\rho_{\mathcal{E}})=\Tr_B(\rho_{\mathcal{E}})=\frac{\mathbb{I}_d}{d}.
\label{LMMcond}
\end{equation}

\section{Constructing locally maximally mixed bipartite states}\label{BipartiteFamily}

We now construct parameterised families of bipartite states with maximally mixed marginals adapting the methodology introduced by Rodriguez-Ramos and Wilmott in \cite{ramos2021convex} for channel construction. Therein unital quantum channels were defined by means of a map sending the elements of a complex matrix into the matrix representation of the quantum channel. Using the same methodology, we construct families of bipartite states which are maximally mixed from a local point of view. This is achieved by defining the set of complex matrices $\mathcal{A}_d:=(\alpha_{ij})_{i,j\in\Z_d}\in\C^{d\times d}$
such that
\begin{equation}
\sum_{i,j=0}^{d-1}{\alpha_{ij}\alpha^*_{ij}}=1
\label{matparam1}
\end{equation}
and
\begin{equation}
\left\{\begin{array}{c}
\sum_{i,j=0}^{d-1}{\alpha_{ij+l}\alpha^*_{ij}}=0 \\ 
\sum_{i,j=0}^{d-1}{\alpha_{ij+l}\alpha^*_{ij}\omega^{-il}}=0
\end{array}\right.
\label{matparam2}
\end{equation}
for $l=1,...,d-1$ with $\omega=\exp{(\ramuno\frac{2\pi}{d})}$. Next, we map the elements of the set of matrices $\mathcal{A}_d$ to the space of bipartite states. Let $A=\{(a_m,b_m)\}_{m\in\Z}$ denote the set of ordered pairs such that $a_m\neq a_{m'}$ and $b_m\neq b_{m'}$ if and only if $m\neq m'$. Consider now the map $\mathcal{P}(a,b):\Z_d \times \Z_d \rightarrow \Z_{d^2}$ given by $\mathcal{P}(a,b)=a+d(b\mod d)$. We can always find a set $S_d=\{A_0,\dots,A_{d-1}\}$ such that $A_i\in\mathcal{A}_d$ and $A_i\neq A_j$ if $i\neq j$ for which 
\begin{equation}
\label{setS}
\bigcup_{i=0}^{r-1}\mathcal{P}(A_i)\equiv\Z_{d^2}.
\end{equation} 
We construct families of bipartite states with maximally mixed marginals with the use of the map defined as follows.
\begin{definition}
\label{defini}
Let $r,d\in\mathbb{N}$ such that $r\geq d$ and let $S_r=\{A_0,\dots,A_{r-1}\}$ such that equation \eqref{setS} is satisfied. Then, we define $\rho_{AB}(\alpha_{ij})\in \mathcal{B}(\mathcal{H}_d\otimes\mathcal{H}_d)$ where $(\alpha_{ij})\in\mathcal{A}_d$ as the following state

\begin{equation}
\rho_{AB}(\alpha_{ij})=\frac{1}{d}\sum_{n=0}^{r-1}\sum_{\substack{(k,h)\in A_n \\ (i,l)\in A_n} }\left(\sum_{j=0}^{d-1}{\alpha_{hj}\omega^{jk}}\right)\left(\sum_{j=0}^{d-1}{\alpha^*_{ij}\omega^{-jl}}\right)\ket{\mathcal{P}(k,k+h)}\bra{\mathcal{P}(l,l+i)}.
\label{Generbip}
\end{equation}
\end{definition}

Definition \ref{defini} establishes a general method to construct parameterised families of bipartite qudit states with maximally mixed marginals and rank $r$. Here, we will consider the particular family of states for which $r=d$. For this particular family, the bipartite states as given by \eqref{Generbip} can be expressed as
\begin{align}
   \rho_{AB}(\alpha_{ij})=\frac{1}{d}\sum_{i,k,l=0}^{d-1}\left(\sum_{j=0}^{d-1}{\alpha_{ij}\omega^{jk}}\right)\left(\sum_{j=0}^{d-1}{\alpha^*_{ij}\omega^{-jl}}\right)\qudit{k+i}\qudit{k}\rqudit{l+i}\rqudit{l}.
    \label{bipstate}
  \end{align}
The states $\rho_{AB}(\alpha_{ij})$ with $(\alpha_{ij})\in\mathcal{A}_d$ are indeed states with maximally mixed marginals to see that we evaluate its partial traces with are given by
\begin{align}
\Tr_A{(\rho_{AB})}&=\frac{1}{d}\Tr_A{\left(\sum_{i,k,l=0}^{d-1}\left(\sum_{j=0}^{d-1}{\alpha_{ij}\omega^{jk}}\right)\left(\sum_{j=0}^{d-1}{\alpha^*_{ij}\omega^{-jl}}\right)\qudit{k+i}\qudit{k}\rqudit{l+i}\rqudit{l}\right)}\nonumber\\
&=\frac{1}{d}\sum_{i,k=0}^{d-1}\left(\sum_{j=0}^{d-1}{\alpha_{ij}\omega^{jk}}\right)\left(\sum_{j=0}^{d-1}{\alpha^*_{ij}\omega^{-jk}}\right)\qudit{k}\rqudit{k}\nonumber\\
&=\frac{1}{d}\sum_{k,l=0}^{d-1}\left(\sum_{i,j=0}^{d-1}{\alpha_{ij+l}\alpha^*_{ij}\omega^{lk}}\right)\qudit{k}\rqudit{k}
\label{step1}
\end{align}
By \eqref{matparam2}, all the elements of the last sum in \eqref{step1} with $l\neq 0$ cancel out and consequently we obtain that
\begin{align}
\Tr_A{(\rho_{AB})}=\frac{1}{d}\sum_{i,k=0}^{d-1}\left(\sum_{j}^{d-1}{\alpha_{ij}\alpha^*_{ij}}\right)\qudit{k}\rqudit{k}
\end{align}
and by \eqref{matparam1} we obtain that
\begin{equation}
\Tr_A{(\rho_{AB})}=\frac{1}{d}\sum_{k=0}^{d-1}\qudit{k}\rqudit{k}.
\end{equation}
We may also evaluate the second partial trace which is given by
\begin{align}
\Tr_B{(\rho_{AB})}&=\frac{1}{d}\Tr_B{\left(\sum_{i,k,l=0}^{d-1}\left(\sum_{j=0}^{d-1}{\alpha_{ij}\omega^{jk}}\right)\left(\sum_{j=0}^{d-1}{\alpha^*_{ij}\omega^{-jl}}\right)\qudit{k+i}\qudit{k}\rqudit{l+i}\rqudit{l}\right)}\nonumber\\
&=\frac{1}{d}\sum_{i,k=0}^{d-1}\left(\sum_{j=0}^{d-1}{\alpha_{ij}\omega^{jk}}\right)\left(\sum_{j=0}^{d-1}{\alpha^*_{ij}\omega^{-jk}}\right)\qudit{k+i}\rqudit{k+i}\nonumber\\
&=\frac{1}{d}\sum_{i,k=0}^{d-1}\left(\sum_{j=0}^{d-1}{\alpha_{ij}\omega^{j(k-i)}}\right)\left(\sum_{j=0}^{d-1}{\alpha^*_{ij}\omega^{-j(k-i)}}\right)\qudit{k}\rqudit{k}\nonumber\\
&=\frac{1}{d}\sum_{k,l=0}^{d-1}\left(\sum_{i,j=0}^{d-1}{\alpha_{ij+l}\alpha^*_{ij}\omega^{l(k-i)}}\right)\qudit{k}\rqudit{k}\nonumber\\
&=\frac{1}{d}\sum_{k,l=0}^{d-1}\left(\sum_{i,j=0}^{d-1}{\alpha_{ij+l}\alpha^*_{ij}\omega^{-il}}\right)\omega^{kl}\qudit{k}\rqudit{k}.
\label{step2}
\end{align}
By \eqref{matparam2}, we get that all the elements of the sum in \eqref{step2} with $l\neq 0$ cancel out and consequently we obtain that \begin{align}
\Tr_B{(\rho_{AB})}=\frac{1}{d}\sum_{i,k=0}^{d-1}\left(\sum_{j}^{d-1}{\alpha_{ij}\alpha^*_{ij}}\right)\qudit{k}\rqudit{k}
\end{align}
and by \eqref{matparam1}, consequently
\begin{equation}
\Tr_B{(\rho_{AB})}=\frac{1}{d}\sum_{k=0}^{d-1}\qudit{k}\rqudit{k}.
\end{equation}

\section{Local unitary classification of $\rho_{AB}(\alpha_{ij})$}\label{LUinvariants}

In this section, we consider the classification of the states $\rho_{AB}(\alpha_{ij})$ as given by \eqref{bipstate} in terms of the parameters $(\alpha_{ij})\in\mathcal{A}_d$. In particular, we consider their classification in terms of their entanglement properties. We identify sets of constants under local unitary (LU) operations acting on $\rho_{AB}$ and consequently determine equivalent classes of states in terms of entanglement.

\subsection{Spectrum of $\rho_{AB}$}

The spectrum of $\rho_{AB}$ is invariant under global unitary operations and consequently, it is also invariant under local unitary (LU) operations. To obtain the explicit set of eigenvalues for $\rho_{AB}$, we apply a unitary conjugation which block-diagonalises the operator as 
\begin{align}
\tau_U(\rho_{AB})
&=U\rho_{AB} U^{\dagger}\nonumber\\
&=\frac{1}{d}\sum_{i=0}^{d-1}\left(\sum_{j=0}^{d-1}{\alpha_{ij}\omega^{jk}}\qudit{i}\qudit{k}\right)\left(\sum_{j,k=0}^{d-1}{\alpha^*_{ij}\omega^{-jk}}\rqudit{i}\rqudit{k}\right).
\end{align} 
The matrix $\tau_U(\rho_{AB})$ is a block diagonal matrix and its blocks $P_0,\dots,P_{d-1}$ are given by
\begin{align}
P_{i}&=\frac{1}{d}\left(\sum_{k=0}^{d-1}\sum_{j=0}^{d-1}{\alpha_{ij}\omega^{jk}}\qudit{k}\right)\left(\sum_{k=0}^{d-1}\sum_{j=0}^{d-1}{\alpha^*_{ij}\omega^{-jk}}\rqudit{k}\right).
\end{align}
The matrices $P_{0},\dots,P_{d-1}$ are rank one, can be expressed as $P_{i}=\ket{p_i}\bra{p_i}$. The norm of each $\ket{p_i}$ is an eigenvalue of $\rho_{AB}$ which is given by
\begin{align}
\braket{p_i|p_i}&=\frac{1}{d}\sum_{k=0}^{d-1}{}\left(\sum_{j=0}^{d-1}{}\alpha_{ij}\omega^{jk}\right)\left(\sum_{j=0}^{d-1}{}\alpha^*_{ij}\omega^{-jk}\right)\nonumber\\
&=\frac{1}{d}\sum_{k,j,l=0}^{d-1}{}\alpha_{i,j+l}\alpha^*_{ij}\omega^{lk}\nonumber\\
&=\frac{1}{d}\sum_{k,j=0}^{d-1}{}\left(\alpha_{ij}\alpha^*_{ij}+\sum_{l=1}^{d-1}\alpha_{i,j+l}\alpha^*_{ij}\omega^{lk}\right)\nonumber\\
&=\sum_{j=0}^{d-1}{}|\alpha_{ij}|^2.
\end{align}
Note that in the last equality, we used the properties of $(\alpha_{ij})\in\mathcal{A}_d$ as given by \eqref{matparam1}. The constants 
\begin{equation}
\kappa^{(1)}_i=\sum_{j=0}^{d-1}{}|\alpha_{ij}|^2
\label{spectrum}
\end{equation}
determine local unitary equivalence classes for the states $\rho_{AB}$.

\subsection{Spectrum of $\rho_{AB}^{T_B}$} 

The spectrum of the partial transpose of the density matrix is invariant under local unitary operations acting on the original state \cite{zhang2013criterion}. The matrix $\rho_{AB}^{T_B}$ is expressed as
\begin{align}
   \rho_{AB}^{T_B}=\frac{1}{d}\sum_{i,k,l=0}^{d-1}\left(\sum_{j=0}^{d-1}{\alpha_{ij}\omega^{jk}}\right)\left(\sum_{j=0}^{d-1}{\alpha^*_{ij}\omega^{-jl}}\right)\qudit{k+i}\qudit{l}\rqudit{l+i}\rqudit{k}.
\end{align}
We can always find a unitary conjugation $\tau_U$ on the partially transposed matrix which block-diagonalises the  matrix $\rho_{AB}^{T_B}$ as
\begin{align}
\tau_U(\rho_{AB}^{T_B})&=U \rho_{AB}^{T_B} U^{-1}\nonumber\\
&=\frac{1}{d}\sum_{i,k,l=0}^{d-1}\left(\sum_{j=0}^{d-1}{\alpha_{ij}\omega^{jk}}\right)\left(\sum_{j=0}^{d-1}{\alpha^*_{ij}\omega^{-jl}}\right)\qudit{k+l+i}\qudit{l}\rqudit{l+k+i}\rqudit{k}.
\end{align}
and the blocks $Q_0,\dots,Q_{d-1}$ are given by
\begin{equation}
Q_i=\frac{1}{d}\sum_{k,l=0}^{d-1}\left(\sum_{j=0}^{d-1}{\alpha_{i-l-k;j}\omega^{jk}}\right)\left(\sum_{j=0}^{d-1}{\alpha^*_{i-l-k;j}\omega^{-jl}}\right)\qudit{l}\rqudit{k}.
\end{equation}
The spectrum of $\rho_{AB}^{T_B}$ is given by the union of all the blocks $Q_i$ and we denote by $\kappa^{(2)}_0,\dots,\kappa^{(2)}_{d^2-1}$ to the set of local unitary invariant scalars given by
\begin{equation}
\kappa^{(2)}_i\in\{\lambda(Q_0)\cup\dots\cup \lambda(Q_{d-1})\},
\label{spectrum}
\end{equation}
where $\lambda(X)$ denotes the set of eigenvalues of the matrix $X$. 

\subsection{Singular values of the correlation matrix of $\rho_{AB}$}
Bipartite density operators $\rho_{AB}\in\mathcal{B}(\mathcal{H}_A\otimes\mathcal{H}_B)$ can be expressed in terms of the Fano decomposition as
\begin{equation}
\rho_{AB}=\frac{1}{d^2}\left(\id_d\otimes\id_d+\sum_{i}^{d^2-1}{}s_i\lambda_i\otimes\id_d+\sum_{i}^{d^2-1} t_i\id_d\otimes\lambda +\sum_{i,j}^{d^2-1} r_{ij} \lambda_i\otimes\lambda_j\right),
\label{fanoform}
\end{equation}
where $s_i=\Tr{(\rho_{AB}\,\lambda_i\otimes\id_d)}$, $t_i=\Tr{(\rho_{AB}\,\id_d\otimes\lambda_i)}$, $r_{ij}=\Tr{(\rho_{AB}\,\lambda_i\otimes\lambda_j)}$ and $\{\lambda_i\}_{i\in{d^2-1}}$ is a base of traceless  matrices. For states with maximally mixed marginals, we have that $\Tr_A(\rho_{\mathcal{AB}})=\Tr_B(\rho_{\mathcal{AB}})=\frac{\mathbb{I}_d}{d}$ where $s_i=0$ and $t_i=0$ as given in \eqref{fanoform}, and $\rho_{AB}$ admits the expression
\begin{equation}
\rho_{AB}=\frac{1}{d^2}\left(\id_d\otimes\id_d+\sum_{i,j}^{d^2-1} r_{ij} \lambda_i\otimes\lambda_j\right).
\end{equation}
The matrix $R=(r_{ij})_{i,j\in\Z_{d^2-1}}$ is called the correlation matrix of $\rho_{AB}$ and it encodes non-local information about the state. The singular values of the correlation matrix of $\rho_{AB}$ are invariant under local unitary operations and we will use them for the entanglement classification of $\rho_{AB}(\alpha_{ij})$ as given by \eqref{bipstate}.

To evaluate the correlation matrix $R$, we select a  particular basis $\lambda_1,\dots,\lambda_{d^2-1}$ in which $R$ is block-diagonal. We define this basis as follows: First, we define the diagonal elements of the basis as 
\begin{align}
\lambda^{(0)}_i=\sum_{m=0}^{d-1}{}\omega^{im}\ket{m}\bra{m}, 
\label{bas0}
\end{align}
where $i=1,\dots,d-1$ and $\omega$ is a $d$th root of unity. Second, we define the off-diagonal elements of the basis by 
\begin{align}
\lambda^{(k)}_i=\ket{i+k}\bra{i}\quad \text{where} \quad i=0,\dots,d-1 \quad \text{and} \quad k=1,\dots,d-1
\label{bask}
\end{align}
For this particular choice of basis, we find that the correlation matrix $R$ of $\rho_{AB}$ can be expressed in block-diagonal form where the blocks $R_0,\dots, R_{d-1}$ are given by
\begin{equation}
R_{0}=(r^{(0)}_{i,j})_{i,j=1,\dots,d-1},
\label{dblock}
\end{equation}
where \begin{align}
r^{(0)}_{i,j}&=\braket{\rho_{AB},\lambda^{(0)}_i\otimes \lambda^{(0)}_j}\nonumber\\
&=\braket{\rho_{AB},\sum_{m=0}^{d-1}{}\omega^{mi}\ket{m}\bra{m}\otimes \sum_{p=0}^{d-1}{}\omega^{pj}\ket{p}\bra{p}}\nonumber\\
&=\sum_{m,p=0}^{d-1}{}\omega^{mi+pj}\braket{\rho_{AB},\ket{m}\bra{m}\otimes {}\ket{p}\bra{p}}\nonumber\\
&=\frac{1}{d}\sum_{m,p=0}^{d-1}{}\omega^{mi+pj}\left(\sum_{s=0}^{d-1}{\alpha_{m-p,s}\omega^{sp}}\right)\left(\sum_{s=0}^{d-1}{\alpha^*_{m-p,s}\omega^{-sp}}\right),
\end{align}
and
\begin{equation}
R_{k}=(r^{(k)}_{i,j})_{i,j=0,\dots,d-1}
\label{odblock}
\end{equation}
where
\begin{align}
r^{(k)}_{i,j}&=\braket{\rho_{AB},\lambda^{(k)}_i\otimes \lambda^{(k)}_j}\nonumber\\
&=\braket{\rho{AB},\ket{i+k}\bra{i}\otimes \ket{j+k}\bra{j}}\nonumber\\
&=\frac{1}{d}\left(\sum_{s=0}^{d-1}{\alpha_{i-j,s}\omega^{sj}}\right)\left(\sum_{s=0}^{d-1}{\alpha^*_{i-j,s}\omega^{-s(j+k)}}\right).
\end{align}

We denote by $\kappa^{(3)}_1,\dots,\kappa^{(3)}_{d^2-1}$ to the singular values of the correlation matrix $R$ of $\rho_{AB}(\alpha_{ij})$ which are given by
\begin{equation}
\kappa^{(3)}_i\in\{\sigma(R_0)\cup\dots\cup \sigma(R_{d-1})\},
\label{spectrum}
\end{equation}
where $\sigma(X)$ denotes the set of singular values of the matrix $X$. We summarize all the local unitary invariant scalars of $\rho_{AB}$ in Table \ref{kappas}.
\begin{table}
\label{table}
\begin{center}
\begin{tabular}{ |c|c|c| }
\hline
Scalar & Definition  & Cardinality \\
\hline
$\kappa^{(1)}_i$ & $\sum_{j=0}^{d-1}{}|\alpha_{ij}|^2$ & $i\in\Z_d$\\
$\kappa^{(2)}_i$ & $\lambda(Q_0)\cup\dots\cup\lambda(Q_{d-1})$ & $i\in\Z_{d^2}$\\
$\kappa^{(3)}_i$ & $\sigma(R_0)\cup\dots\cup \sigma(R_{d-1})$ & $i\in\Z_{d^2-1}$\\
\hline
\end{tabular}
\end{center}
\caption{In this table we summarize the sets of scalars which are invariant under local unitary operations $\rho_{AB}$ used for entanglement classification.}
\label{kappas}
\end{table}

The values of $\kappa^{(i)}$ are related to well-known quantum state measures. For example, $\kappa^{(1)}_1,\kappa^{(1)}_2$ and $\kappa^{(1)}_3$, derived from the spectrum of the density matrix, are associated with the purity of $\rho_{AB}$. The purity of a density matrix is defined as $\mathcal{P}(\rho_{AB})=\Tr(\rho^2_{AB})$. Equivalently we can express the purity of $\rho_{AB}$ in terms of $\kappa^{(1)}_i$ by
\begin{equation}
\mathcal{P}(\rho_{AB})=\sum_{i=1}^d\left(\kappa^{(1)}_i\right)^2.
\label{purity}
\end{equation}
Similarly, the sets $\kappa^{(2)}_i$ and $\kappa^{(3)}_i$ are related to other entanglement measures for $\rho_{AB}$. In particular, the set $\kappa^{(2)}_i$ relates to the negativity of the quantum state which is defined in terms of the partially transposed state $\rho^{T_A}$ as 
\begin{equation}
\mathcal{N}(\rho_{AB})=\frac{||\rho^{T_A}||_1-1}{2},
\end{equation}
where $||X||_1$ denotes the trace norm of the operator $X$. Equally, we can express the negativity in terms of the eigenvalues of $\rho_{AB}^{T_A}$ and consequently, we can express the negativity of $\rho_{AB}$ in terms of $\kappa^{(2)}_i$ as \begin{equation}
\mathcal{N}(\rho_{AB})=\sum_{i=1}^{d^2}\frac{|\kappa^{(2)}_i|-1}{2}.
\label{negativity}
\end{equation}
Finally set $\kappa^{(3)}_i$ relates to the quantum discord of the density state \cite{luo_geometric_2010}.

\section{Qutrit case}\label{QutritCase}

In section \ref{LUinvariants}, we obtained sets of local unitary invariant scalars for the family of bipartite states in \eqref{Generbip}. Now, we consider the particular case of qutrit states. For $d=3$, the local unitary invariant scalars of $\rho_{AB}$ are given by 
\begin{align}
\kappa^{(1)}_i=\sum_{j=0}^{2}{}\alpha_{ij}\alpha^*_{ij},
\end{align} for $i=0,1,2$. We can also obtain $\kappa^{(2)}_1,\dots,\kappa^{(2)}_9$ which are given by the eigenvalues of the matrices  
\begin{align}
Q_{0}&=\begin{pmatrix}
c_{0,0}c^*_{0,0}& c_{2,0}c^*_{2,1}& c_{1,0}c^*_{1,2}\\
c_{2,1}c^*_{2,0}& c_{1,1}c^*_{1,1}& c_{0,1}c^*_{0,2}\\
c_{1,2}c^*_{1,0}& c_{0,2}c^*_{0,1}& c_{2,2}c^*_{2,2}
\end{pmatrix},\nonumber\\
Q_{1}&=\begin{pmatrix}
c_{1,0}c^*_{1,2}& c_{0,1}c^*_{0,0}& c_{2,2}c^*_{2,1}\\
c_{0,0}c^*_{0,2}& c_{2,1}c^*_{2,0}& c_{1,2}c^*_{1,1}\\
c_{2,0}c^*_{2,2}& c_{1,1}c^*_{1,0}& c_{0,2}c^*_{0,1}
\end{pmatrix},\nonumber\\
Q_{2}&=\begin{pmatrix}
c_{2,0}c^*_{2,2}& c_{1,1}c^*_{1,0}& c_{0,2}c^*_{0,1}\\
c_{1,0}c^*_{1,2}& c_{0,1}c^*_{0,0}& c_{2,2}c^*_{2,1}\\
c_{0,0}c^*_{0,2}& c_{2,1}c^*_{2,0}& c_{1,2}c^*_{1,1}
\end{pmatrix};
\end{align}
where $c_{i,k}=\frac{1}{d}\left(\sum_{j=0}^{d-1}{\alpha_{ij}\omega^{jk}}\right)$. For qutrits, the LU invariant scalars $\kappa^{(3)}_1,\dots,\kappa^{(3)}_8$ are given by the singular values of the matrices
\begingroup
\allowdisplaybreaks
\begin{align}
R_{0}&=\sum_{m,p=0}^{2}{}c_{m-p,p}c^*_{m-p,p}\begin{pmatrix}
\omega^{m+p}&\omega^{2m+p}\\
\omega^{m+2p}&\omega^{2m+2p}
\end{pmatrix},\nonumber\\
R_{1}
&=\begin{pmatrix}
c_{0,0}c^*_{0,1}& c_{2,1}c^*_{2,2}& c_{1,2}c^*_{1,0}\\
c_{1,0}c^*_{1,1}& c_{0,1}c^*_{0,2}& c_{2,2}c^*_{2,0}\\
c_{2,0}c^*_{2,1}& c_{1,1}c^*_{1,2}& c_{0,2}c^*_{0,0}
\end{pmatrix},\nonumber\\
R_{2}
&=\begin{pmatrix}
c_{0,0}c^*_{0,2}& c_{2,1}c^*_{2,0}& c_{1,2}c^*_{1,1}\\
c_{1,0}c^*_{1,2}& c_{0,1}c^*_{0,0}& c_{2,2}c^*_{2,1}\\
c_{2,0}c^*_{2,2}& c_{1,1}c^*_{1,0}& c_{0,2}c^*_{0,1}
\end{pmatrix}.
\end{align}
\endgroup
We note that the evaluation of $\kappa^{(1)}_i$,$\kappa^{(2)}_i$,$\kappa^{(3)}_i$ depends on the specific choice of coefficients $\{\alpha_{ij}\}_{i,j\in\Z_3}$, and we present an example of a bi-parametric family of bipartite states for which we obtain a complete LU classification. 
\subsection{Example: A 2-parameter family of bipartite states}
\begin{figure}[h]
    \centering
    \includegraphics[width=0.7\textwidth]{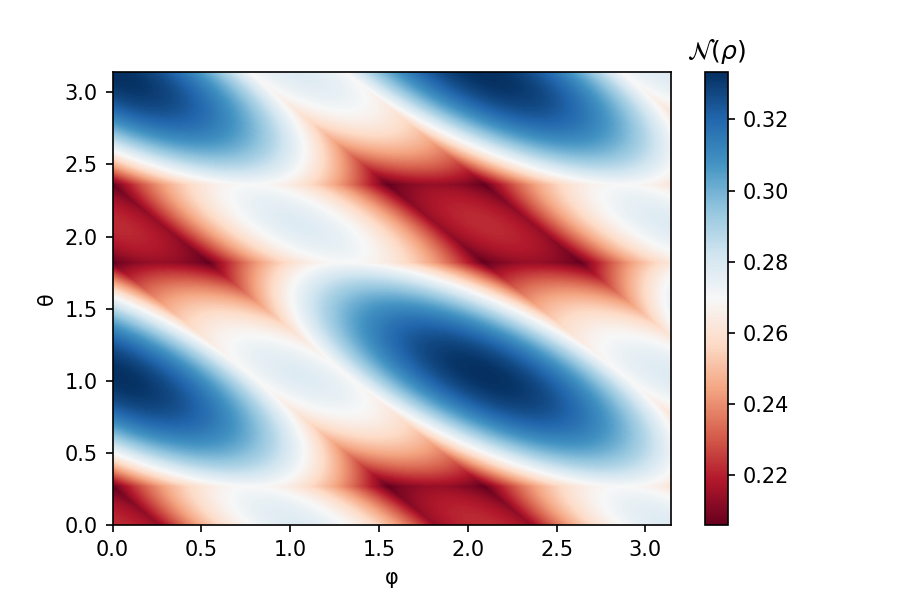}
    \caption{Color plot of $\mathcal{N}(\rho_{AB})$ in terms of the two parameters spanning the family of states represented by \eqref{2family}. }
    \label{fig:mesh1}
\end{figure}

The set of complex matrices given by
\begin{equation}
(\alpha_{ij})_{i,j\in\Z_3}=
\frac{\sqrt{2}}{6}\begin{pmatrix}
2 & \euler^{\ramuno\theta} & \euler^{\ramuno(\theta+\phi)}\\
2 & \euler^{\ramuno\theta-\frac{2\pi}{3}} & \euler^{-\ramuno(\theta+\phi-\frac{4\pi}{3})}\\
2 & \euler^{\ramuno\theta-\frac{4\pi}{3}} & \euler^{-\ramuno(\theta+\phi-\frac{2\pi}{3})}\\
\end{pmatrix}
\label{2family}
\end{equation}
determines a 2-dimensional family of bipartite qutrit states with maximally mixed marginals. We can check the matrices given in \eqref{2family} satisfy $(alpha_{ij})_{i,j\in\Z}\in{\mathcal{A}_3}$ and consequently, this set of matrices can be mapped to a family of bipartite states with maximally mixed marginals. For this family of states, we might evaluate the set of local unitary invariant scalars $\kappa^{(1)},\kappa^{(2)},\kappa^{(3)}$ derived in the previous section. For this particular family, the first set of LU invariant scalars corresponds to the spectrum of the density matrix. These scalars are given by
\begin{equation}
\kappa^{(1)}_1=\kappa^{(1)}_2=\kappa^{(1)}_3=\frac{1}{3}.
\end{equation}
By equation \eqref{purity}, we evaluate the purity of the family of states given by \eqref{2family} as 
\begin{equation}
\mathcal{P}(\rho_{AB})=\frac{1}{3}.
\end{equation}
For this particular family, the set $\kappa^{(2)}_i$ consists of nine scalars given by the spectrum of the partially transposed matrix. We have it that
\begin{align}
\kappa^{(2)}_{1+i}&=\frac{1}{18}\left(4+2\cos(\phi+\frac{2i\pi}{3})\right)\nonumber\\
\kappa^{(2)}_{4+i}&=\frac{1}{18}\left(1+4\cos(\theta+\frac{2i\pi}{3})\right)\nonumber\\
\kappa^{(2)}_{7+i}&=\frac{1}{18}\left(1+4\cos(\theta+\phi+\frac{4i\pi}{3})\right)
\end{align} for $i=0,1,2$. By equation \eqref{negativity} we can evaluate the negativity of the family of states represented by \eqref{2family}. Figure \ref{fig:mesh1} represents the negativity of this family of states as a function of the parameters $\theta$ and $\phi$. We observe that negativity is upper bounded by $\frac{1}{3}$.

Finally, we can evaluate the set of LU-invariant scalars  set $\kappa^{(3)}_i$ corresponding to the singular values of the correlation matrix. For the particular family of states represented by \eqref{2family}, we obtain the following list of scalars
\begin{align}
\kappa^{(3)}_{1+i}&=\frac{1}{6}
\nonumber\\
\kappa^{(3)}_{3+i}&=\frac{1}{18}\sqrt{9-4\cos{(\theta-\phi)}-8\cos{(-2\theta-\phi)}-4\cos{(\theta+2\phi)}}
\nonumber\\
\kappa^{(3)}_{5+i}&=\frac{1}{18}\sqrt{9-4\cos{(\theta-\phi+\frac{\pi}{3})}-8\cos{(-2\theta-\phi+\frac{\pi}{3})}-4\cos{(\theta+2\phi+\frac{\pi}{3})}}
\nonumber\\
\kappa^{(3)}_{7+i}&=\frac{1}{18}\sqrt{9-4\cos{(\theta-\phi-\frac{\pi}{3})}-8\cos{(-2\theta-\phi-\frac{\pi}{3})}-4\cos{(\theta+2\phi-\frac{\pi}{3})}},
\end{align}
for $i=0,1$. For this particular family of states, the entanglement classification provided by $\kappa^{(1)}_i$, $\kappa^{(2)}_i$ and $\kappa^{(3)}_i$ is complete.

\section{Conclusion}

 In this work, we considered the entanglement classification of bipartite states with maximally mixed marginals. First, we constructed a parameterized family of bipartite states which generalises the anti-symmetric Werner states. Second, we obtained a set of scalars which remain constant under local unitary operations and, consequently, can be used for entanglement classification. In particular, we used the eigenvalues density matrix and the partially transposed matrix as well as the singular values of the correlation matrix to classify the elements of our family of bipartite states. Finally, we considered the qutrit scenario which is the smallest set-up for which a complete entanglement classification is missing. For qutrit bipartite states, we evaluated the sets of local invariant scalars and we relate them to known measures of entanglement. As an example, we showed that this set achieves a complete classification of a bi-parameter family of qutrit states. We believe that this work serves as an intermediate step to achieve better quantum state classification.

\printbibliography

\end{document}